\documentstyle[12pt,epsf]{article}
\advance \topmargin by -\headheight
\advance \topmargin by -\headsep
\evensidemargin \oddsidemargin
\begin{document}
\begin{titlepage}
\section*{The role of the quantum dispersion in the Coulomb correction 
             of Bose-Einstein correlations}

\begin{center}

\vspace{0.5cm}

H. Merlitz$^{a,b}$ and D. Pelte$^{a,c}$\\
\end{center}

\vspace{0.5cm}

  $a$: Physikalisches Institut der Universit\"at Heidelberg,
  D-69120 Heidelberg\\

  $b$: Gesellschaft f\"ur Schwerionenforschung (GSI) Darmstadt, Germany\\

  $c$: Max-Planck-Institut f\"ur Kernphysik, D-69117 Heidelberg\\

\begin{abstract}
 The time dependent Schr\"odinger equation for two 
 identical and charged pions is solved using 
 wavepacket states. It is shown that the expected
 Coulomb distortion in the momentum correlation function
 is obliterated by the dispersion of the localized states,
 and therefore becomes unobservable.\\
 PACS: 25.70.-z\\
 Keywords: HBT-Analysis; Coulomb correction; Wavepackets\\
 E-mail: pelte@pel5.mpi-hd.mpg.de
\end{abstract}
\end{titlepage}

\section{Introduction}
 \label{intro}

 The Bose-Einstein correlations of identical charged pions 
 are believed to be distorted by the Coulomb interaction
 between both pions. Since the end of the seventies, several efforts 
 where made to investigate these distortions. A reliable 
 information about the size of the pion source can only
 be extracted from the correlation signal when the properties
 of the Coulomb distortions are sufficiently understood.
 The size of the Coulomb effect, and how it should be calculated,
 was always a source of intensive
 discussions. In the early theoretical approaches, the square
 of the relative Coulomb wave function for a pointlike source, the
 Gamow factor, was employed to obtain the Coulomb correction
 of the correlation function \cite{gyu79}. Later it was realized that
 this ansatz overestimated the Coulomb suppression, and since then
 different models where developed which successively led to smaller and
 smaller distortions. First, the finite source size was included
 \cite{bow91}, later the influence of the exchange terms 
 \cite{biy95},
 the influence of long-lived resonances \cite{sin96}, of Coulomb
 screening due to other charged particles \cite{bay96} and the
 finite momentum resolution of the detector \cite{bar97}. 
 
 Already in 1986, however, it became clear that a correct treatment
 of pions in the fireball must include their localization in
 terms of quantum mechanical wavepackets, the spatial widths of which
 are connected to the mean free path of the pions which has
 a value of a few fm \cite{pra86}. 
 Recently, a consistent wavepacket description was presented by the
 authors and applied to different numerical codes 
 \cite{mer97}. A somewhat different ansatz, which may be
 suitable especially for very high multiplicities,
 is developed in \cite{cso97}. It was demonstrated, how the 
 localization of the pions introduces a
 zero point energy, which consequently leads to a momentum uncertainty
 in the pion ensemble, i.\,e.\ to dispersion. In order to estimate
 the possible influence of the dispersion onto the Coulomb effects,
 we compare the size of the Coulomb potential with the  
 zero point energy using a simple
 picture: Imagine two pions which are separated by the distance 
 $r_{\rm o}$. Both states shall not overlap significantly, so that we
 require them to be localized in Gaussian wavepackets with the spatial
 extension $\sigma_{\rm o} = r_{\rm o}/2$. We evaluate the Coulomb
 energy $E_{\rm coul} = e^2/r_{\rm o}$
 and the zero point (quantum-) energy
  $E_{\rm q} = 3\hbar^2/(8m\sigma_{\rm o}^2)$
 for a distance of $r_{\rm o} = 4$ fm and obtain $E_{\rm coul}
 \approx 0.4$ MeV and $E_{\rm q} \approx 26$ MeV.
 Therefore, the quantum energy $E_{\rm q}$ is nearly two
 orders of magnitude larger than $E_{\rm coul}$. 
 Only for distances $r_{\rm o} >
 290$ fm exceeds the Coulomb energy the localization energy
 and the picture of two classical particles with Coulomb repulsion  
 becomes valid. For smaller distances the dispersion introduces
 a strong random motion into the ensemble which obliterates the
 directed motion caused by the Coulomb repulsion. 
 This suggests that a correct quantum mechanical treatment
 of the Coulomb interaction is necessary and can not be substituted
 by any semiclassical model. 

\section{Matrix representation of the Schr\"odinger equation}
 \label{matrix}
 The Schr\"odinger equation for two identical charged bosons,
 \begin{equation}
  \label{eq30}
  \hat{H} |\Psi({\bf r}_1, {\bf r}_2,t)\rangle =
  i\hbar \partial_t |\Psi({\bf r}_1, {\bf r}_2,t)\rangle\;,
 \end{equation}
 with the Hamiltonian
 \begin{equation}
  \label{eq40}
   \hat{H} = -\frac{\hbar^2}{2m}(\Delta_1 + \Delta_2) + \frac{e^2}{
   |{\bf r}_1 - {\bf r}_2|^2} \equiv \hat{T} + \hat{V}\
 \end{equation}
 and the symmetrized 2- particle state 
 \begin{equation}
  \label{eq50}
  |\Psi({\bf r}_1, {\bf r}_2,t)\rangle =
  {\cal N}_2\left(|\chi_1({\bf r}_1,t)\rangle \,|\chi_2({\bf r}_2,t)\rangle +
  |\chi_2({\bf r}_1,t)\rangle\,|\chi_1({\bf r}_2,t)\rangle\right)\;,
 \end{equation}
 has to be solved. Since the Hamiltonian does not depend on time, the
 solution could be found on principle by time independent
 methods. The problem, however, are the initial conditions of
 the system: At $t=0$ each particle is
 localized within a certain volume (wavepackets in the fireball). 
 There exists neither an analytical solution for this problem nor
 a way how to transform these initial conditions into appropriate
 border conditions for a time independent treatment. We therefore
 have to solve the time dependent problem. 
 To start with, we expand the state (\ref{eq50}) into a set of basis functions
 \begin{equation}
  \label{eq60}
  |\Psi({\bf r}_1, {\bf r}_2,t)\rangle =
  \sum c_k(t) \, |\Psi_k({\bf r}_1, {\bf r}_2,t)\rangle
 \end{equation}
 with the c-valued coefficients 
 \begin{equation}
  \label{eq70}
  c_k(t) = \langle \Psi_k({\bf r}_1, {\bf r}_2,t)|\Psi({\bf r}_1,
               {\bf r}_2,t)\rangle\;.
 \end{equation}
 Now the Schr\"odinger equation can be written in matrix form 
 \begin{equation}
  \label{eq110}
  i\hbar\, {\bf \dot{c}} = \tilde{S}^{-1} \, \left(\tilde{H} - 
  \tilde{D}\right)\, {\bf c}
 \end{equation}
 with the time dependent coefficient vector ${\bf c}$, the
 Hamilton matrix 
 $\tilde{H}_{ij} = \langle \Psi_i | \hat{T} + \hat{V} | \Psi_j \rangle$,
 the overlap matrix
 $\tilde{S}_{ij} = \langle \Psi_i | \Psi_j \rangle$ and the
 time evolution matrix
 $\tilde{D}_{ij} = \langle \Psi_i | i\hbar \partial_t | \Psi_j \rangle$.  
 This representation of the Schr\"odinger equation is equivalent
 to Eq.\ (\ref{eq30}) if the basis provides a complete set  
 of states in the 2- particle Hilbert space. In any numerical
 calculations, of course, only a limited number of basis functions
 can be used so that an adequate choice of the basis is
 essential in order to reach the required accuracy and, at the same
 time, to keep the numerics within affordable limits.   

\section{Choice of the basis set}
 \label{basis} 
 For the single particle states we use the following time dependent
 wave functions:
 \begin{equation}
   \label{eq120}
   \chi_{\vec{l}_jk}({\bf r},t) =
   x_{k}^{l_{jx}}\, y_{k}^{l_{jy}}\, z_{k}^{l_{jz}}
   \exp\left(
   \frac{i}{\hbar} \left({\bf P}_k(t)\cdot
   {\bf r} - \theta_{\vec{l}_j,k}(t) \right) -
   \frac{\left({\bf r} - {\bf R}_k(t)\right)^2}
        {4s(t)\sigma_{\rm o}}\right)
 \end{equation}
 with the phase
 \begin{equation}
  \label{eq130}
  \theta_{\vec{l}_j,k}(t) = \int_{0}^{t}E_k^{\rm kin}(\tau) \, d\tau +
  \hbar\, (l_{jx} + l_{jy} + l_{jz} + 3/2)\,
  \arctan\left(\frac{\hbar t}{2m\sigma_o^2}\right)
 \end{equation}
 and $x_{k}^{l_{jx}} = ({\bf r}_x - {\bf R}_{k_x})^{l_{jx}}$, and similar for
 the $y$- and $z$ coordinates. The capital coordinates 
 $({\bf R}_k(t),{\bf P}_k(t))$ denote the phase space coordinates of the
 $k$'th wave packet centre, whereas ${\bf r}$ is the free variable of
 the state in configuration space representation. The time dependent width
 is given by $s(t) = \sigma_{\rm o} + i \hbar t/(2 m \sigma_{\rm o})$
 and the initial width is $\sigma_{\rm o}$. With the index vector
 $\vec{l}_j = (0,0,0)$ we
 get the usual Gaussian wave packet which was already used in \cite{mer97}.
 Following the notation used in the harmonic oscillator problem, this function
 may be called ``s- type'' function, and consequently the index vector
 $\vec{l}_j = (1,0,0)$ defines a ``p- type'' function and so on.   
 We now build symmetrized 2- particle states from the single particle
 states, because then the basis functions already fulfill the correct 
 permutation symmetry. An additional symmetry requirement is 
 that the squared amplitude must be symmetric with respect 
 to the centre of mass of the system, which requires the index
 vectors of both single particle states to be equal.
 We therefore define the basis in terms of the homogeneous states
 \begin{equation}
  \label{eq180}
  \Psi_{\vec{l}_i} \sim
            \chi_{\vec{l}_i1}({\bf r}_1,t)\,\chi_{\vec{l}_i2}({\bf r}_2,t) +
            \chi_{\vec{l}_i2}({\bf r}_1,t)\,\chi_{\vec{l}_i1}({\bf r}_2,t)\;.
 \end{equation}
 We use a set of 2- particle basis functions, which 
 contains 1 s- function, 3 p- functions, 6 d- functions and 10 f- functions.  
 The use of even larger basis sets is prohibited by the 
 exponentially increasing numerical effort. For example,
 the evaluation of a complete trajectory (which 
 typically requires $500$ timesteps) needs $1$ minute 
 for a sp- basis, 1/2 hour for a spd- basis and 1/2 day
 for a spdf- basis\footnote{Values obtained on a DEC 
 $250^{4/266}$ workstation.}.  

\section{The solution of Schr\"odinger's equation}
 \label{solution}
 In order to solve the Schr\"odinger equation (\ref{eq110}),
 the matrices $\tilde{V}$, $\tilde{T}$, $\tilde{D}$ and $\tilde{S}$
 have to be evaluated. It is possible to obtain
 analytical expressions for the kinetic energy- and time
 evolution matrix elements, whereas the potential matrix elements,
 representing a 6- fold integral, can be reduced to an one dimensional
 integration and tabulated. Details of the very lengthy calculations
 may not be given here. We refer to the literature in quantum 
 chemistry\footnote{For example, in \cite{cle89}, Chapter 6, the
 calculations are done for the time independent Gaussians. For
 the evaluation of our time dependent basis set, a detailed
 script (13 pages) can be obtained from the authors.}. 

 At the beginning of each event, the system has to be initialized. 
 As source function we used the following simple parametrization of the phase
 space:
 \begin{equation}
 \label{eq200}
  \rho ({\bf R})=\frac{1}{(\pi R_s^2)^{\,\frac{3}{2}}}
  \exp\left( \frac{-{\bf R}^2}{R_s^2} \right)\;\;\;\mbox{and}\;\;\;\;\;
   f({\bf P}) = (2\pi m T)^{-3/2}\, \exp\left(\frac{-{\bf P}^{2}}{2 m
   T}\right)\;,
 \end{equation}
 i.\,e.\ the source is a static Gaussian with rms- radius $\sqrt{3/2}R_s$
 and temperature $T$. This choice is motivated by the fact that 
 an analytical expression for the 2- particle correlation function is
 available for noninteracting wavepackets
 \cite{mer97}:
 \begin{equation}
  \label{eq220}
  {\cal C}_{2}({\bf p}_1, {\bf p}_2) = \sum_{k=0}^{\infty} (-1)^k\,
  \left(
  \frac{a_o^2 b_o^2}{a_k b_k c_k d_k}\right)^{3/2}
  \,\left(e^{f_k({\bf p}_1, {\bf p}_2)} + e^{g_k({\bf p}_1, {\bf p}_2)}\right)
 \end{equation}
 with
 \begin{eqnarray}
  \label{eq230}
  f_k({\bf p}_1, {\bf p}_2) &=&
  - ({\bf p}_1 - {\bf p}_2)^2 \frac{k}{16\sigma_p^6 a_k c_k}
  \;,\\
  g_k({\bf p}_1, {\bf p}_2) &=&
  - ({\bf p}_1 - {\bf p}_2)^2 \left(
  \frac{4 \sigma_p^2 a_k + k}{32 \sigma_p^6 a_k c_k} +
  \frac{1}{2\hbar^2 R_s^2 b_k d_k}\right)  \;,
  \label{eq240}
 \end{eqnarray}
 and
 \begin{eqnarray}
  \nonumber
  a_k &=& \frac{1}{2\sigma_p^2} + \frac{1}{2mT} + \frac{k}{4\sigma_p^2}\;,
  \;\;\;\;\; b_k = \frac{1}{R_s^2} + \frac{k}{4\sigma_o^2}\;,\\
  c_k &=& a_k - \frac{k^2}{16\sigma_p^4 a_k}\;,\;\;\;\;\;
  d_k = b_k - \frac{k^2}{16\sigma_o^4 b_k}\;,
  \label{eq250}
 \end{eqnarray}
 $\sigma_{\rm p} = \hbar/(2\sigma_{\rm o})$ is the width of
 the wavepacket in momentum space. 
 The state was initialized to be a pure s- state, i.\,e.\ 
 with the coefficient vector ${\bf c}(t=0) = (1,0,\dots,0)$
 in Eq.\ (\ref{eq110}),
 and the initial width of the Gaussian was $\sigma_{\rm o} = 1.8$ fm. The time
 propagation was performed using a second order differential scheme
 \cite{lef91}. For each timestep, the force between both wavepackets was
 obtained numerically using $\dot{{\bf P}}_2 \equiv -\nabla_{21} 
 \langle \Psi |\hat{V}({\bf R}_{21})| \Psi \rangle$ and
 $\dot{{\bf P}}_1 = - \dot{{\bf P}}_2$. Here, ${\bf R}_{21}$ is the difference
 vector between both centres. The propagation was stopped when the residual
 Coulomb energy had dropped below $5$\% of the initial value. 
 At the end of each event, the final state was Fourier transformed into
 the momentum space representation. Such a state
 defines a quantum mechanical (pure) ensemble of equally prepared systems.
 We can extract an arbitrary number of representatives by using the squared
 amplitude as density function and applying the Monte Carlo 
 procedure presented in \cite{mer97} to sample out a number
 of $N_s$ correlated pairs. The 2- particle correlation function, however,
 is built by the mixed ensemble which is obtained by averaging over the
 source function. Therefore we have to calculate a large number $N_e$ of
 events with different initial conditions according to the source densities
 Eq.\ (\ref{eq200}), yielding $N_e\cdot N_s$ pairs
 which finally obey the correct mixed statistics.
 
\section{Results}
 \label{results}
 The source parameters were set to $R_s = 5$ fm and $T = 50$ MeV.
 In order to gain an intuitive feeling about the strength of the Coulomb effect,
 ``pions'' with different charges $Z = 1$, $Z = 3$ and $Z = 5$ were 
 simulated. Since the Coulomb potential between both particles increases
 quadratically with $Z$, this set covers a range between the realistic 
 case and the case where the Coulomb energy nearly reaches the same 
 magnitude as
 the localization energy as was discussed in Sect.\ \ref{intro}. 
 $N_e = 2000$ 
 events were generated using a sp- basis set, the average system 
 propagation time was $250$ fm/c until the Coulomb potential had
 decayed below $5$\% of its initial value. Table \ref{tab1} displays
 some key parameters for the different systems. Clearly the average
 initial Coulomb energy increases quadratically with the particle
 charge (second column). The third column demonstrates that around
 70\% of this initial Coulomb energy is converted into kinetic energy
 of the wavepackets, independent of the charge. This implies that
 the residual 30\% are put into excitation of the p- basis
 functions, i.\,e.\ into deformation of the wavepackets. Since with
 the sp- basis we only have $4$ basis functions which are 
 orthogonal, we can use the fact that 
 \begin{equation}
  \label{eq310}
  |c_s| + |c_{p_x}| + |c_{p_y}| + |c_{p_z}| = 1
 \end{equation}
 to compare the weights that are present in the final state.
 The fourth column shows that in case of $Z = 1$ in the average only
 0.4\% of the weight is deposited into the p- basis, whereas 99.6\%
 remain in the s- function. This implies that the state remains
 nearly undisturbed. In case of $Z = 5$ the situation is different:
 Already 8\% of the weight is put into the p-   
 functions, in 10\% of the events this weight even exceeds 20\%.
 A considerable deformation is therefore induced by the Coulomb force.

\begin{figure}
  \epsfxsize=12.0cm
  \epsffile{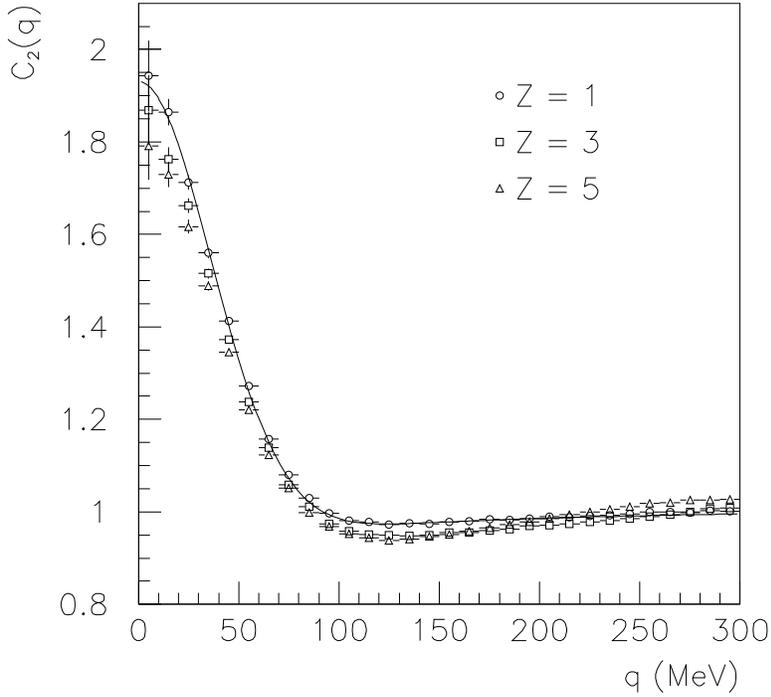}
  \caption{\label{fig1} 
  The 2- particle correlation functions
  for different charges. The solid curve is the theoretical
  result for noninteracting pions Eq.\ (\protect\ref{eq220}).}
 \end{figure}

 For the evaluation of the 2- particle correlation function, each
 of the $2000$ generated 
 events was used to extract $N_s = 5000$ correlated pairs, yielding
 a statistics of $10^7$ pairs. Figure \ref{fig1} displays the resulting
 correlation functions. Two facts are of importance:
 \begin{enumerate}
  \item For $Z = 1$ the distortion of the correlation function is invisible.
  \item For $Z = 3$ and $Z = 5$ an increasing deviation from the 
        undistorted correlation appears, but the effect is global
        in the sense that the whole correlation function is involved
        rather than only the part of low momentum differences.
 \end{enumerate}
 Several test simulations were performed with different basis sets
 including sp-, spd- and spdf- sets in order to check whether the
 used sp- basis is already sufficiently large. For the $Z = 1$ simulation no
 differences were observed, which implies that the sp-
 basis is sufficient to yield reliable results. This is not surprising
 since the deformation of the wavepackets is very small as discussed
 above. For the $Z = 5$ simulation some dependence on the basis size
 could be observed. The size of these effects suggests that a
 simulation with spdf- basis instead the sp- basis may lead to
 further changes of the correlation function, which may
 be as large as 20\% of the distortions calculated with the 
 smaller basis. Such simulations, however, could not be performed
 with sufficient statistics due to the huge numerical effort.

 As a second time dependent method, the molecular dynamics procedure
 which was discussed in \cite{mer97} and is based on Bohm's quantum
 theory of motion, was employed to evaluate the
 pion-pion Coulomb interaction. In this method, only the Gaussian
 s- functions are used. The additional degree of freedom is included
 by means of an equation of motion for certain test particles,
 which contains the quantum potential and
 accounts for dispersion- and correlation
 effects. Since it is the distribution of the test particles
 which is used to deduce the correlation function, the deformation
 of the wavepacket is achieved indirectly due to their Coulomb-
 distorted positions. This ansatz is approximate in the sense
 that the distorted test particles lead to distorted density
 distributions and therefore to a distorted state, a fact that
 in an exact treatment would change the quantum forces and therefore
 the particle trajectories and again (in higher order) the state.
 But for only small distortions it was found to be sufficient
 to evaluate the quantum force using the undisturbed state.

\begin{figure}
  \epsfxsize=12.0cm
  \epsffile{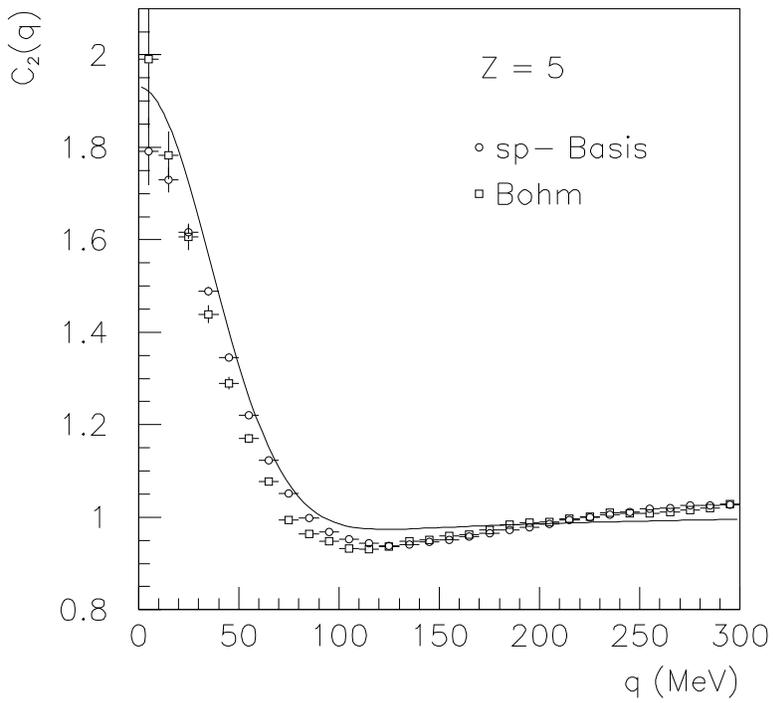}
  \caption{\label{fig2}  
  Correlation function for $Z = 5$
  and two different numerical methods. The solid curve is 
  the theoretical result for noninteracting pions 
  Eq.\ (\protect\ref{eq220}).}
 \end{figure}

 The simulations were repeated for the $Z = 1$ and $Z = 5$ case. 
 A number of
 $3\cdot 10^6$ pion pairs were generated for each charge
 and used to determine the correlation functions. For $Z = 1$ we
 obtained the same result as with the sp- basis expansion, namely
 no visible Coulomb distortion. Figure \ref{fig2} only displays
 the result of the $Z = 5$ simulation in comparison to the 
 corresponding result from the basis expansion. An overall
 qualitative agreement is observed, for small momentum differences,
 however, there exist some deviations. This indicates the possible
 inaccuracy of both methods, since for the 
 hypothetical and rather extreme case
 $Z = 5$ the Coulomb effects are no longer small and both methods
 begin to suffer from the approximative character of the small basis
 respective the unproper evaluation of the quantum force. For the 
 $Z = 1$ simulation, on the other hand, the approximations
 are well justified and the calculations suggest that the pion-pion
 Coulomb interaction has no visible effect onto the 2- particle 
 correlation function and therefore experimental data
 must not be ``corrected'' to remove a nonexistent effect.
 
\section{Why scattering theory fails}
 \label{fail}
 Usually, the pion-pion Coulomb effects are treated by
 considering the time independent scattering of
 plane waves in a Coulomb potential. The shortcomings
 of this ansatz are most obvious if we write the correct
 quantum mechanical ansatz in the Bohm formalism. We may
 initialize a wavepacket of width $\sigma_{\rm o}$ 
 with the centre coordinate ${\bf R}_{\rm o}$ and a
 test particle at the position ${\bf x}_{\rm o}$
 and obtain the following equation of motion for the 
 test particle:  
 \begin{equation}
  \label{eq320}
   m\frac{d^2{\bf x}}{dt^2} = -\nabla(V + Q)\;
 \end{equation}
 with the quantum potential
 \begin{equation}
  \label{eq330}
   Q(t = 0) = \frac{\hbar^2}{4m\sigma_{\rm o}^2}\,
   \left(3 - \frac{({\bf x}_{\rm o}  - {\bf R}_{\rm o})^2}
   {2\sigma_{\rm o}^2}\right)\;,
 \end{equation}
 which exerts a repulsive force onto the test particle and
 which when integrated over the density distribution yields the
 zero point energy of the wavepacket \cite{mer97}. In the
 plane wave limit ($\sigma_{\rm o}\rightarrow \infty$), the
 quantum potential vanishes and (\ref{eq320}) becomes a
 classical equation of motion. This is the reason why the
 results from scattering theory can be made to agree 
 with pure classical trajectory calculations \cite{bay96}.  
 As a crosscheck, we did repeat the $Z = 1$ simulation for
 wavepackets without dispersion, i.\,e.\ for a fixed width. The 
 resulting correlation function is in 
 agreement with the predictions of the classical ``toy model'' 
 \cite{bay96}, which proves that it is the dispersion which
 masks the Coulomb effect.


 In conclusion, we have shown that the correlation functions
 in pion interferometry are not distorted by pion-pion
 Coulomb effects. Due to the initial conditions, which force
 the pions to be localized within a small volume of a few fm$^3$,
 the zero point energy produces a dispersion which makes the
 comparably tiny distortion due to Coulomb interaction
 to become unobservable. 
 For heavier particles the strength of the dispersion
 decreases and therefore then the Coulomb effect is expected
 to be of more relevance. We have
 supported our argumentation by numerical simulations using two
 different methods. In the first (basis expansion), the Coulomb
 influence is accounted for on the level of the state amplitude,
 in the other (Bohm's molecular dynamics)
 on the level of the equation of motion. Scattering 
 theory fails to produce the correct results, because with the
 removal of the quantum potential in Eq.\ (\ref{eq320})  
 the most important part of the dynamical evolution is 
 neglected. 
 This implies that experimental data, which are published
 after ``Coulomb correction'', are wrong for small momentum 
 differences.

 \newpage
 \pagestyle{empty}

 \begin{table}
   \caption{\label{tab1} 
   Averaged values for the initial
   Coulomb energy (2'nd column), the difference between
   final and initial kinetic energy (3'rd column)
   and the degree of excitation of the p- basis
   coefficients (4'th column) for different charges.
   The numbers (in brackets)
   mean that in 10\% of the events that value was
   exceeded.}
   \begin{center}
     \begin{tabular}[t]{|c||c|c|c|}  \hline
     $ Z $ & $\langle E_{\rm coul} \rangle$ (MeV)
           & $\langle \Delta E_{\rm kin}\rangle$   (MeV)
           & $\langle 1 - |c_s| \rangle$\\
     \hline
     1   & 0.21 (0.35)& 0.15 (0.25)& 0.004 (0.01)\\
     3   & 1.9 (3.2)  & 1.3 (2.5)  & 0.03 (0.09) \\
     5   & 5.2 (9.0)  & 3.7 (7.0)  & 0.08 (0.2)  \\ \hline
     \end{tabular}
   \end{center}
 \end{table}

 \newpage
 \pagestyle{empty}

 \newpage
 \pagestyle{empty}

\end{document}